\RequirePackage{lineno} 
\documentclass{nature}
\usepackage{xcolor}

\usepackage{caption}
\usepackage{multirow}
\usepackage{amsmath}
\usepackage{bm}
\usepackage{braket}
\usepackage{amsmath}
\usepackage{upgreek}

\title{Experimental evidence of plasmarons and effective fine structure constant in electron-doped graphene/h-BN heterostructure}

\author{Hongyun Zhang$^1$, Shuopei Wang$^2$, Eryin Wang$^1$, Xiaobo Lu$^2$, Qian Li$^1$, Changhua Bao$^1$, Ke Deng$^1$, Haoxiong Zhang$^1$, Wei Yao$^{1}$, Guorui Chen$^3$, Alexei V. Fedorov$^4$, Jonathan D. Denlinger$^4$, Kenji Watanabe$^5$, Takashi Taniguchi$^6$, Guangyu Zhang$^2$ and Shuyun Zhou$^{1,7,*}$}

\usepackage{graphicx}
\makeatletter
\let\saved@includegraphics\includegraphics
\AtBeginDocument{\let\includegraphics\saved@includegraphics}

\makeatother
\begin{document}
%\linenumbers
\maketitle

\begin{affiliations}
 \item State Key Laboratory of Low Dimensional Quantum Physics and Department of Physics, Tsinghua University, Beijing 100084, P.R. China.
\item Beijing National Laboratory for condensed Matter Physics and Institute of physics, Chinese Academy of Science, Beijing 100190, P.R. China.
\item State Key Laboratory of Surface Physics and Department of Physics, Fudan University, Shanghai 200433, China.
\item Advanced Light source, Lawrence Berkeley National Laboratory, Berkeley, California 94720, USA.
\item Research Center for Functional Materials, National Institute for Materials Science, 1-1 Namiki, Tsukuba 305-0044, Japan.
\item International Center for Materials Nanoarchitectonics, National Institute for Materials Science, 1-1 Namiki, Tsukuba 305-0044, Japan.
\item Frontier Science Center for Quantum Information, Beijing 100084, P. R. China. \\

* Correspondence should be sent to syzhou@mail.tsinghua.edu.cn.
\end{affiliations}

%\newpage

\begin{abstract}
	
{\bf Electron-electron interaction is fundamental in condensed matter physics and can lead to composite quasiparticles called plasmarons, which strongly renormalize the dispersion and carry information of electron-electron coupling strength as defined by the effective fine structure constant $\alpha_{ee}^*$. Although h-BN with unique dielectric properties has been widely used as an important substrate for graphene, so far there is no experimental report of plasmarons in graphene/h-BN yet. Here, we report direct experimental observation of plasmaron dispersion in graphene/h-BN heterostructures through angle-resolved photoemission spectroscopy (ARPES) measurements upon {\it in situ} electron doping. Characteristic diamond-shaped dispersion is observed near the Dirac cone in both 0$^\circ$ (aligned) and 13.5$^\circ$ (twisted)  graphene/h-BN, and the electron-electron interaction strength $\alpha_{ee}^*$ is extracted to be $\alpha_{ee}^*\approx0.9\pm 0.1$, highlighting the important role of electron-electron interaction. Our results suggest graphene/h-BN as an ideal platform for investigating strong electron-electron interaction with weak dielectric screening, and lays fundamental physics for gate-tunable nano-electronics and nano-plasmonics.	
}

%(-$\infty$,-$\Omega_1$)    (-$\Omega_1$, -$\Omega_2$)     (-$\Omega_2$,0)
\end{abstract}

\newpage

\section*{Introduction}

Electron-electron interaction is ubiquitous in solids and plays an important role in condensed matter physics. In graphene, the strength of the electron-electron interaction is quantified by the ratio of the Coulomb potential $U= \rm e^2k_F/\epsilon$ to the kinetic energy $K=\hbar {\rm v_F} \rm k_F$,\cite{RevModPhys_2009,Antonio_RevModPhys2012,NatPhotonics_2012} where ${\rm v_F}$ and $\rm k_F$ are the Fermi velocity and Fermi momentum respectively, and  $\epsilon$ is the dielectric constant.  This ratio defines a fundamental constant $\alpha_{ee}^*=U/K = {\rm e^2}/\epsilon \hbar {\rm v_F}$,\cite{AbbamonteSci, Nair_Sciencec2008} which is in analogy to the fine structure constant $\alpha = {\rm e^2}/\hbar c = 1/137$ in quantum electrodynamics, with the speed of light $c$ replaced by the Fermi velocity ${\rm v_F}$ and the effective dielectric screening of the environment taken into account by $\epsilon$.  Because of this analogy, $\alpha_{ee}^*$ is called the effective fine structure constant. The effective fine structure constant $\alpha_{ee}^*$ reflects the relative strength of electron-electron interaction and determines many fundamental physical properties of graphene, e.g., optical absorption\cite{Nair_Sciencec2008} and transport properties.\cite{FuhrerPRL2008} Moreover, by tuning the carrier concentration and electron-electron correlation, superconductivity has been reported in twisted bilayer graphene placed on h-BN substrate.\cite{PabloMott2018} Revealing the electron-electron interaction strength and extracting the effective fine structure constant are fundamentally important. 

Electron-electron interaction can significantly affect the electronic dispersion by reshaping the graphene Dirac cone dispersion with a modified Fermi velocity.\cite{LanzaraPRL2013,LanzaraPNAS,GeimReshaped,KatochPRB2020} Moreover, collective excitations of electron gas can form plasmons,\cite{PinesPR1953} and the interaction between plasmons and charges leads to composite quasiparticles called plasmarons,\cite{Lundqvist_1967} which strongly modify the electronic dispersion with newly generated plasmaron bands. As an atomically thin material with unique conical dispersion as shown in Fig.~1a, graphene shows a stronger plasmon-charge interaction\cite{Hwang_PRB2007,PoliniPRB2008,HwangPRB2008} than other two-dimensional (2D) materials with parabolic dispersions in general. In particular, the coupling between charges and plasmons with the same group velocity leads to  low-energy plasmaron dispersions displaced from the original Dirac cone, forming a characteristic diamond-shaped dispersion near the Dirac point\cite{Hwang_PRB2007,PoliniPRB2008,HwangPRB2008,Eliplasmaron2010,EliPRB2011} as schematically illustrated in Fig.~1b.  Since the energy and momentum separation between the Dirac cone and plasmaron bands is determined by the effective fine structure constant $\alpha_{ee}^*$,\cite{Hwang_PRB2007,PoliniPRB2008,HwangPRB2008,Eliplasmaron2010,EliPRB2011} observing plasmaron dispersion allows to extract $\alpha_{ee}^*$ experimentally, which is critical for revealing the fundamental physics of electron-electron interaction in graphene-based electronics and plasmonics. 

Plasmarons have been reported for graphene grown on SiC substrates by angle-resolved photoemission spectroscopy (ARPES) measurements,\cite{Eliplasmaron2010,EliPRB2011} however, for graphene/h-BN heterostructure which shows  intriguing physics such as Hofstadter butterfly states and newly generated Dirac cones,\cite{LeRoyNP2012,GeimNat2013,Dean2013,HuntSci2013,Yu_2014,EryinNaturephysics,LeRoy_Rev2019} better graphene electronics\cite{Hone2010} and plasmonics\cite{KoppensPlasmon2015a,BasovPlasmon2015, Sunku_Science2018} performance, so far, there is still no experimental report of the plasmarons yet. Here we provide direct experimental evidence for plasmon-charge interaction induced plasmarons in a highly electron-doped graphene/h-BN heterostructure through ARPES measurements combined with {\it in situ} electron doping via surface deposition of Rb (Fig.~1c). Plasmaron bands are clearly observed at high electron doping, forming a diamond-shaped dispersion near the Dirac point. Moreover, from the plasmaron bands, the extracted effective fine structure constant of $\alpha_{ee}^*\approx0.9\pm 0.1$ is found to be the largest among all reported graphene systems including graphite\cite{AbbamonteSci,Abbamonte_PRB2016} and graphene on different treated SiC substrates,\cite{Eliplasmaron2010,EliPRB2011} indicating the strongest electron-electron interaction in graphene/h-BN heterostructure with the weakest effective dielectric screening. In addition, despite the strong plasmaron, the dispersions remain sharp at high electron doping. This suggests the plasmarons and dopants do not significantly affect the electron scattering at such high electron doping, which is useful for gate-tunable nano-electronics and nano-plasmonics applications.

\begin{figure*}%[htbp]
	{\label{fig1a}}
	{\label{fig1b}}
	{\label{fig1c}}
	\centering
	\includegraphics[width=17cm]{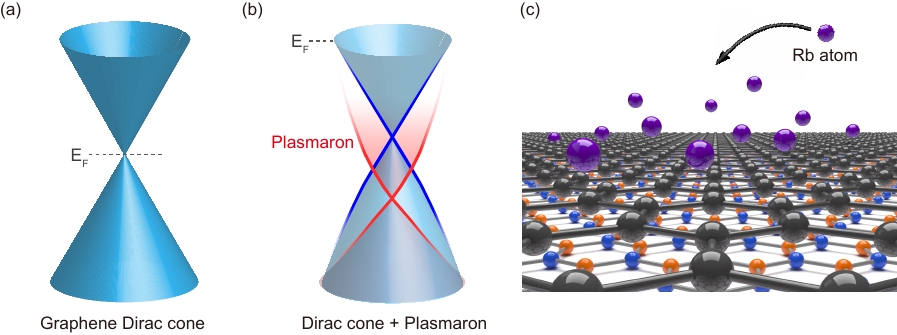}
	\caption{{\bf A schematic drawing for plasmarons in highly electron-doped graphene/h-BN heterostructure.} (a) Dirac cone dispersion in undoped sample. (b) Plasmaron bands plus Dirac cone in highly electron-doped graphene/h-BN. Blue and red cones represent the graphene Dirac cone and plasmaron bands respectively. (c) A schematic drawing of electron doping of graphene/h-BN heterostructure by {\it in situ} Rb deposition.} 
	\label{fig1}
\end{figure*}

\section*{Results}
\section*{Electron doping of graphene/h-BN and gap filling by Rb deposition}

Figure~2 shows an overview of the band structure evolution through the K point (indicated by dotted line in Fig.~2i) upon electron doping.  In the undoped sample, the graphene Dirac point energy ${\rm E_0}$ is near the Fermi energy ${\rm E_F}$ as pointed by blue arrow in Fig.~2a, and the energy of second generation Dirac cones (minibands) ${\rm E_{SDC}}$ is at $\approx$ -200 meV as indicated by gray arrow.  Gap openings  consistent with previous ARPES measurements\cite{EryinNaturephysics} show up as the suppression of intensity near ${\rm E_0}$ and ${\rm E_{SDC}}$.  
Since pristine graphene/h-BN is almost charge neutral with ${\rm E_0}$ near the Fermi energy (${\rm E_F}$), increasing the carrier concentration is important for observing the effects of plasmons which require a dense electron gas.\cite{RevModPhys_2009,PhysRevB_1992}  By {\it in situ} deposition of Rb on the graphene/h-BN surface, we can control the charge carrier concentration in a large range up to 5.8$\times$10$^{13}$ cm$^{-2}$ (calculated from the size of the Fermi pockets;\cite{Luttinger_1960} see  Supplementary Figure 1) and reveal the evolution of the electronic structure. 
Upon electron doping, both ${\rm E_0}$ and ${\rm E_{SDC}}$ shift down in energy while at the same time, gradual filling of the intensity is also observed in the gap region near ${\rm E_0}$ and ${\rm E_{SDC}}$ in Fig.~2b-d (see Supplementary Figure 2 for more details). At carrier concentration of 8.1$\times$10$^{12}$ cm$^{-2}$ (Fig.~2d), the intensity suppression near ${\rm E_0}$ and ${\rm E_{SDC}}$ becomes almost undetectable, indicating gap filling induced by the electronic screening of the inversion-asymmetric component of the moir\'e superlattice potential. Further electron doping leads to the emergence of plasmaron features, which are the main focus of this letter. 

\begin{figure*}%[htbp]
	{\label{fig2a}}
	{\label{fig2b}}
	{\label{fig2c}}
	{\label{fig2d}}
	{\label{fig2e}}
	{\label{fig2f}}
	{\label{fig2g}}
	{\label{fig2h}}
	{\label{fig2i}}
	{\label{fig2j}}	
	\centering
	\includegraphics[width=17cm]{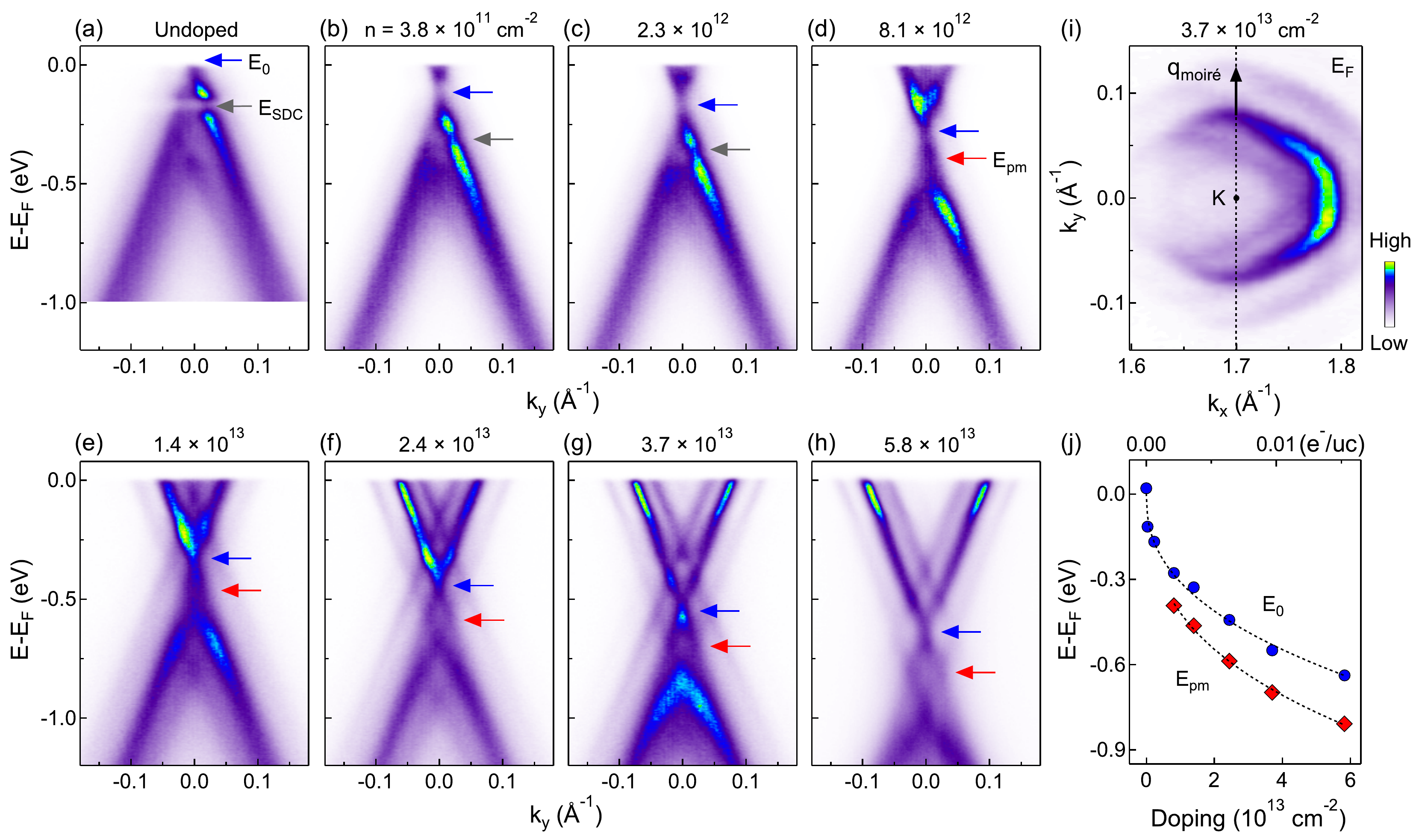}
	\label{Fig.2|}
	\caption{ {\bf Formation of plasmaron bands upon doping in a 0$^\circ$ aligned graphene/h-BN.} (a)-(h) Evolution of the electronic structure through the K point (indicated by black dotted line in i) upon electron doping. ${\rm E_0}$, ${\rm E_{SDC}}$ and ${\rm E_{pm}}$ indicate the Dirac point energy, second-generation Dirac point energy, and plasmaron crossing energy respectively. (i) Fermi surface map at doping level corresponding to data shown in (g). (j) Dirac point energy ${\rm E_0}$ and plasmaron crossing energy ${\rm E_{pm}}$ as a function of electron doping. Black broken curves are guides to the eyes.}
	\label{fig2}
\end{figure*}

\section*{Emergence of plasmaron bands in $0^{\circ}$ aligned graphene/h-BN upon electron doping}

Signatures of plasmaron bands first become observable at a carrier concentration of 8.1$\times$10$^{12}$ cm$^{-2}$ as two twisted bands near the Dirac point (marked by red arrows in Fig.~2d,e). The twisted bands become more obvious and increase in size with doping, and they eventually evolve into a clear diamond-shaped dispersion at even higher electron doping (Fig.~2f-2h), indicating the emergence of plasmaron bands.  This is different from graphene on SiC substrate where no diamond-shaped dispersion is observed\cite{Zhou2007}.
In addition to the diamond-shaped dispersion near the Dirac point energy, moir\'e replica bands are also clearly observed on both sides of the graphene bands near ${\rm E_F}$ (Fig.~2e-2h). Such moir\'e superlattice replicas are also observed as replica pockets in the Fermi surface map shown in Fig.~2i. To summarize the evolution of the electronic structure upon electron doping, the Dirac point energy ${\rm E_0}$ and plasmaron crossing energy ${\rm E_{pm}}$ (defined as the crossing energy between graphene Dirac cone and the plasmaron bands, indicated by red arrows in Fig.~2d-2h) are both plotted in Fig.~2j. Plasmaron bands are clearly observed at intermediate to high electron doping with Dirac point energy ${\rm E_0}$ shifted from -0.28 eV to -0.64 eV (blue symbols in Fig.~2j). Such shift in ${\rm E_0}$ corresponds to carrier concentration ranging from 8.1$\times$10$^{12}$ cm$^{-2}$ to 5.8$\times$10$^{13}$ cm$^{-2}$.  With increasing electron doping, the energy separation between ${\rm E_0}$ and ${\rm E_{pm}}$ becomes larger. We note that the moir\'e superlattice period is determined by the lattice mismatch between graphene and h-BN\cite{YangNaturematerials2013} and is doping independent, and as a result, the separation between the moir\'e superlattice replica and the original Dirac cone is expected to be doping independent. Therefore, the increasing energy and momentum separation between the plasmaron bands and Dirac cone upon electron doping  (see Supplementary Figure 3 for more details) confirms that it is not caused by overlapping of graphene Dirac cone and moir\'e superlattice bands, but instead band renormalization induced by plasmon-charge interaction. In addition, the ARPES dispersions remain quite sharp, and indeed they are sharper than those at lower carrier concentration. This is in agreement with previous report on graphene/h-BN at a lower doping (with the Dirac point at $\approx$ -0.3 eV),\cite{LanzaraPRL2013} where the decrease of scattering rate is attributed to the increase of long-range impurity screening from the higher electron density.  Here we show that at an even higher doping (with the Dirac point at -0.64 eV) and in the presence of plasmarons, the electron scattering rate still remains low, which is useful for gate-tunable nano-electronics and plasmonics.

\begin{figure*}%[tbhp]
	{\label{fig3a}}
	{\label{fig3b}}
	{\label{fig3c}}
	{\label{fig3d}}
	{\label{fig3e}}
	{\label{fig3f}}
	{\label{fig3g}}
	{\label{fig3h}}
	{\label{fig3i}}
	{\label{fig3j}}
	\centering
	\includegraphics[width=18cm]{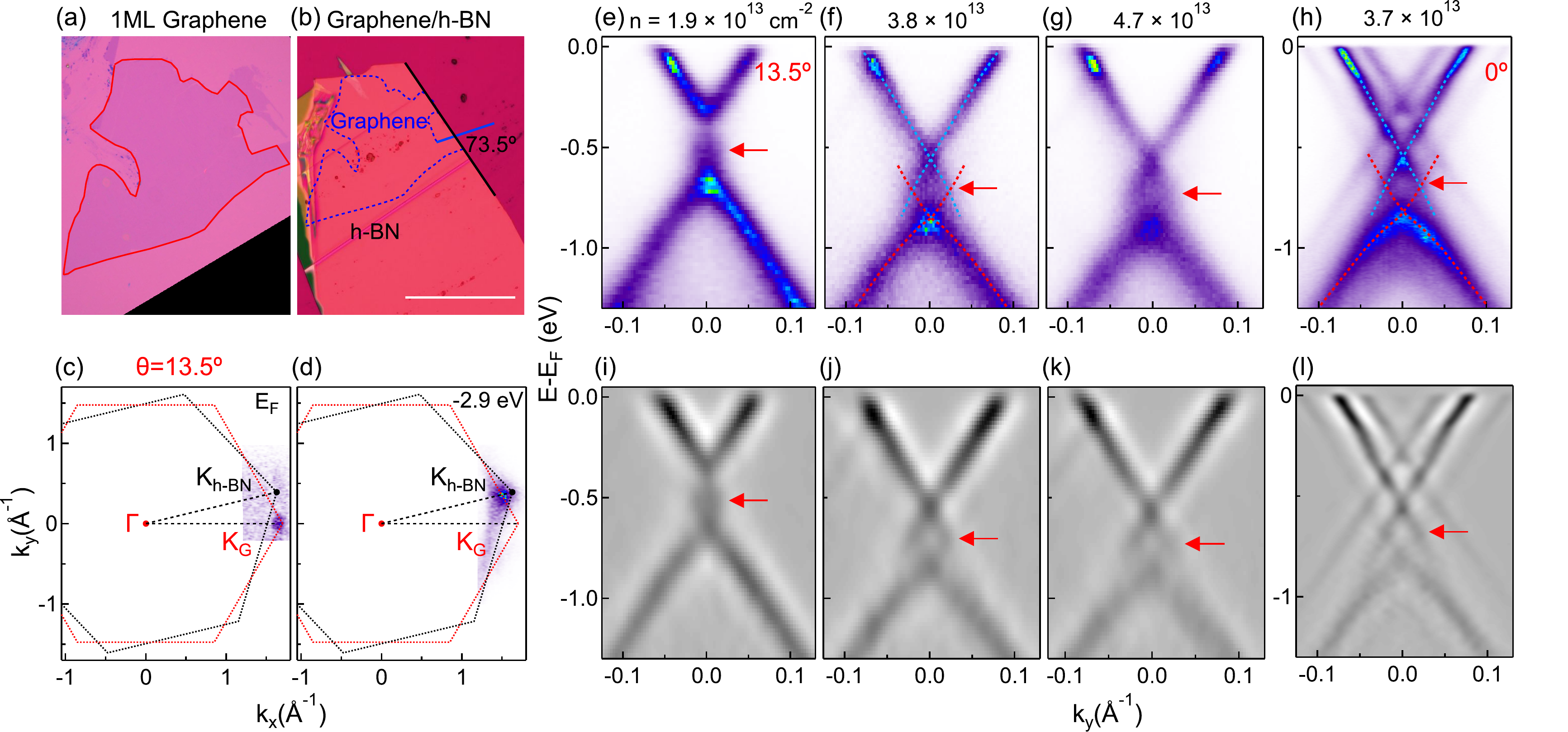}
	\caption{{\bf Plasmaron dispersion observed in a $13.5^{\circ}$ twisted graphene/h-BN heterostructure.} (a,b) Optical images of exfoliated monolayer graphene and graphene/h-BN heterostructure. The twist angle determined from the straight edge is $73.5^{\circ}$ (equivalent to $13.5^{\circ}$). Scale bar indicates 100 $\upmu$m. (c,d) Energy contours at Fermi energy and -2.9 eV to reveal the K points of graphene and h-BN, which confirm the twist angle of $13.5^{\circ}$. (e)-(g) Dispersion of $13.5^{\circ}$ graphene/h-BN at different carrier density to show the evolution of plasmaron bands upon doping. (h) Dispersion of $0^{\circ}$ graphene/h-BN at doping level of 3.7$\times$10$^{13}$ cm$^{-2}$ (also shown in Fig.~2(g)) to compare with results on $13.5^{\circ}$ graphene/h-BN sample. (i)-(l) 2D curvature results of (e)-(h), which clearly show the diamond-shaped dispersion (indicated by red arrows) formed by the Dirac cone and plasmaron bands.}
	\label{fig3}
\end{figure*}

\section*{Plasmaron bands observed in a $13.5^{\circ}$ twisted graphene/h-BN upon electron doping}
Since the moir\'e superlattice period of graphene/h-BN strongly depends on the twist angle\cite{LeRoyNP2012}, to check whether the plasmaron features depend on the twist angle, we show in Fig.~3 ARPES results on a $13.5^{\circ}$ twisted graphene/h-BN heterostructure. The moir\'e superlattice period decreases from $\lambda$ $\approx$ 14 nm at $0^{\circ}$ to 1.05 nm at $13.5^{\circ}$, and therefore the superlattice replica Dirac cone is much farther away from the graphene Dirac cone. Figure~3a shows the exfoliated monolayer graphene, which was then transferred onto a h-BN flake (shown in Fig.~3b) with a designed twist angle of $13.5^{\circ}$. This twist angle is confirmed by the K points of graphene and h-BN from energy contours at the Fermi energy (Fig.~3c) and -2.9 eV (Fig.~3d). By performing ARPES measurements with {\it in situ} Rb deposition, the evolution of the band dispersion upon electron doping is revealed. Figure~3e-3g shows the dispersions measured at different carrier density, with a characteristic diamond-shaped dispersion formed by the crossing of Dirac cone (blue curve) and plasmaron bands (red). The plasmaron bands are more clearly observed in the 2D curvature plots shown in Fig.~3i-3k.  A comparison of dispersions measured at high electron doping for 13.5$^\circ$ twisted (Fig.~3f) and 0$^\circ$ aligned  (Fig.~3h) graphene/h-BN heterostructures shows that the diamond-shaped dispersion has the same size despite the different twist angle. We further note that the Rb doping does not induce any additional reconstruction, as revealed by the Fermi surface map in the doped sample as shown in Supplementary Figure 4. Therefore, the diamond-shaped dispersion near the Dirac point is a result of the plasmaron at high electron doping, independent of the stacking angle between monolayer graphene and h-BN and the distribution of Rb atoms. In the case of magic-angle twisted bilayer graphene where the flat band emerges\cite{MacDonaldPNAS,PabloMott2018,PabloSC2018}, the plasmon dispersion will depend on the twisting angle between the two graphene layers, leading to flat plasmons with a much larger fine structure constant\cite{PlasmonPNAS19} with $\alpha \gg 1$ and chiral edge plasmons\cite{PlasmonTBLGPRL2020}.

\begin{figure*}%[htbp]
	{\label{fig4a}}
	{\label{fig4b}}
	{\label{fig4c}}
	{\label{fig4d}}
	{\label{fig4e}}
	{\label{fig4f}}
	{\label{fig4g}}
	{\label{fig4h}}
	{\label{fig4i}}
	\centering
	\includegraphics[width=17cm]{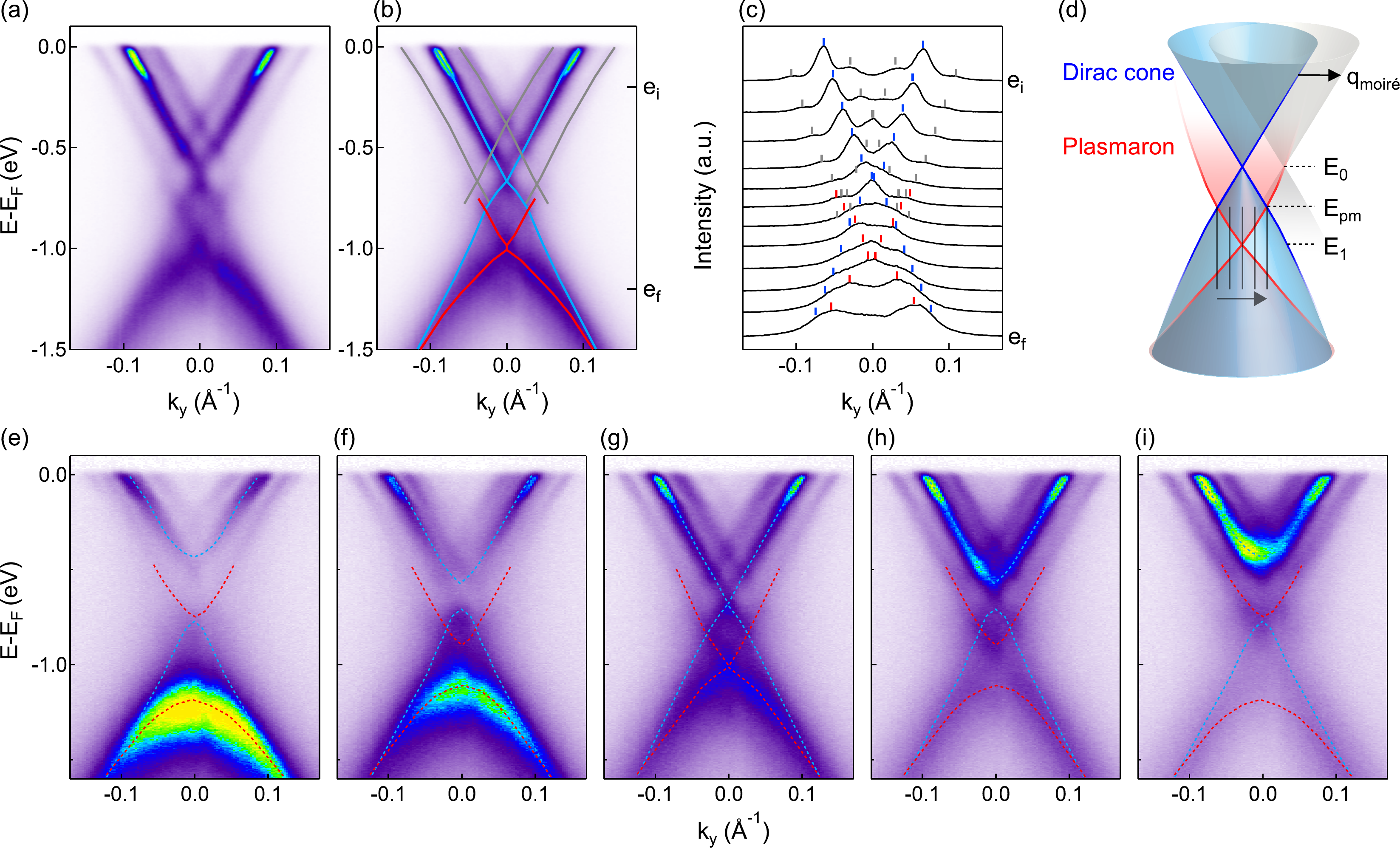}
	\caption{{\bf Electronic structure at doping level of 5.8$\times$10$^{13}$ cm$^{-2}$ to show the plasmaron dispersion.} (a,b) Electronic structure at the highest doping (5.8$\times$10$^{13}$ cm$^{-2}$) through the K point along the K-K direction. Dispersions extracted from MDCs are appended as blue, red and gray curves in (b). (c) MDCs with energy range from $e_i$ to $e_f$ shown in b, and the colored marks indicate peaks and are over-plotted in (b). (d)  A schematic of electronic structure at the highest doping level. Blue, red and gray colors represent the original graphene Dirac cone, plasmaron bands and the moir\'e superlattice bands, respectively. (e)-(i) Cuts parallel to the K-K direction throughout the momentum range of the diamond shape (indicated by gray lines in (d)).}
	\label{fig4}
\end{figure*}

\section*{The plasmaron dispersion at the highest electron doping}

To extract the plasmaron dispersion, we show in Fig.~4 the band structure analysis at the highest doping of 5.8$\times$10$^{13}$ cm$^{-2}$ on $0^{\circ}$ aligned graphene/h-BN sample. Figure~4a shows dispersion image through the Dirac point. Dispersions extracted from peak positions (indicated by tick marks in Fig.~4c in the momentum distribution curves (MDCs) are overplotted in the dispersion image in Fig.~4b. Three types of dispersing bands are identified, the graphene Dirac cone (blue), the plasmaron bands (red) and the moir\'e superlattice bands (gray). The clear diamond-shaped dispersion formed by the crossing of the Dirac cone (blue) and the plasmaron bands (red) is observed in our data in Fig.~4c and is the characteristic feature of plasmarons.  
Figure~4d shows a schematic summary of the band structure.  To follow the evolution of the dispersion across the K point, parallel cuts from one side of the diamond-shaped dispersion to the other side (indicated by gray lines in Fig.~4d) are shown in Fig.~4e-4i.  Both the Dirac cone (indicated by blue dotted curves) and the plasmaron bands (indicated by red dotted curves) show a separation-touching-separation behavior between the conduction and valence bands across this momentum region, confirming the diamond-shaped dispersion along both $k_x$ and $k_y$ direction consistent with the schematic drawing in Fig.~4d. We note that the upper plasmaron bands are much weaker compared to the Dirac cone due to disorder-induced damping\cite{PoliniPRB2008,HwangPRB2008} and near ${\rm E_F}$, the dispersion is dominated by the moir\'e replica bands (gray curves). In addition, as shown in Fig.~3f,h, the diamond-shaped dispersion formed by the Dirac cone and plasmaron is observed in both aligned and twisted graphene/h-BN heterostructures with the same size, indicating the same strength of electron-plasmon interaction.

\section*{The extracted effective fine structure constant}

The plasmaron dispersion in graphene/h-BN heterostructure provides critical information about the electron-electron interaction and the effective fine structure constant.  Although observations of plasmons and plasmarons require large electron concentration, the dimensionless effective fine structure constant $\alpha_{ee}^*$ = ${\rm e^2}$/$\epsilon \hbar {\rm v_F}$ is directly related to the effective dielectric constant $\epsilon$ and is independent of the carrier density. 
In particular, as a result of the unique linear dispersion, the normalized energy separation $\delta E$ (by the separation between the Dirac point energy and the Fermi energy) and momentum separations $\delta k$ (by Fermi momentum $\rm k_F$) between the graphene Dirac cone and the plasmaron dispersion is determined by $\alpha_{ee}^*$.\cite{PoliniPRB2008,Eliplasmaron2010} 
Therefore, the dimensionless effective fine structure constant $\alpha_{ee}^*$ can be directly extracted by analyzing the diamond-shaped dispersion.

\begin{figure*}%[tbhp]
	{\label{fig5a}}
	{\label{fig5b}}
	{\label{fig5c}}
	{\label{fig5d}}
	{\label{fig5e}}
	{\label{fig5f}}
	{\label{fig5g}}
	{\label{fig5h}}
	{\label{fig5i}}
	\centering
	\includegraphics[width=17cm]{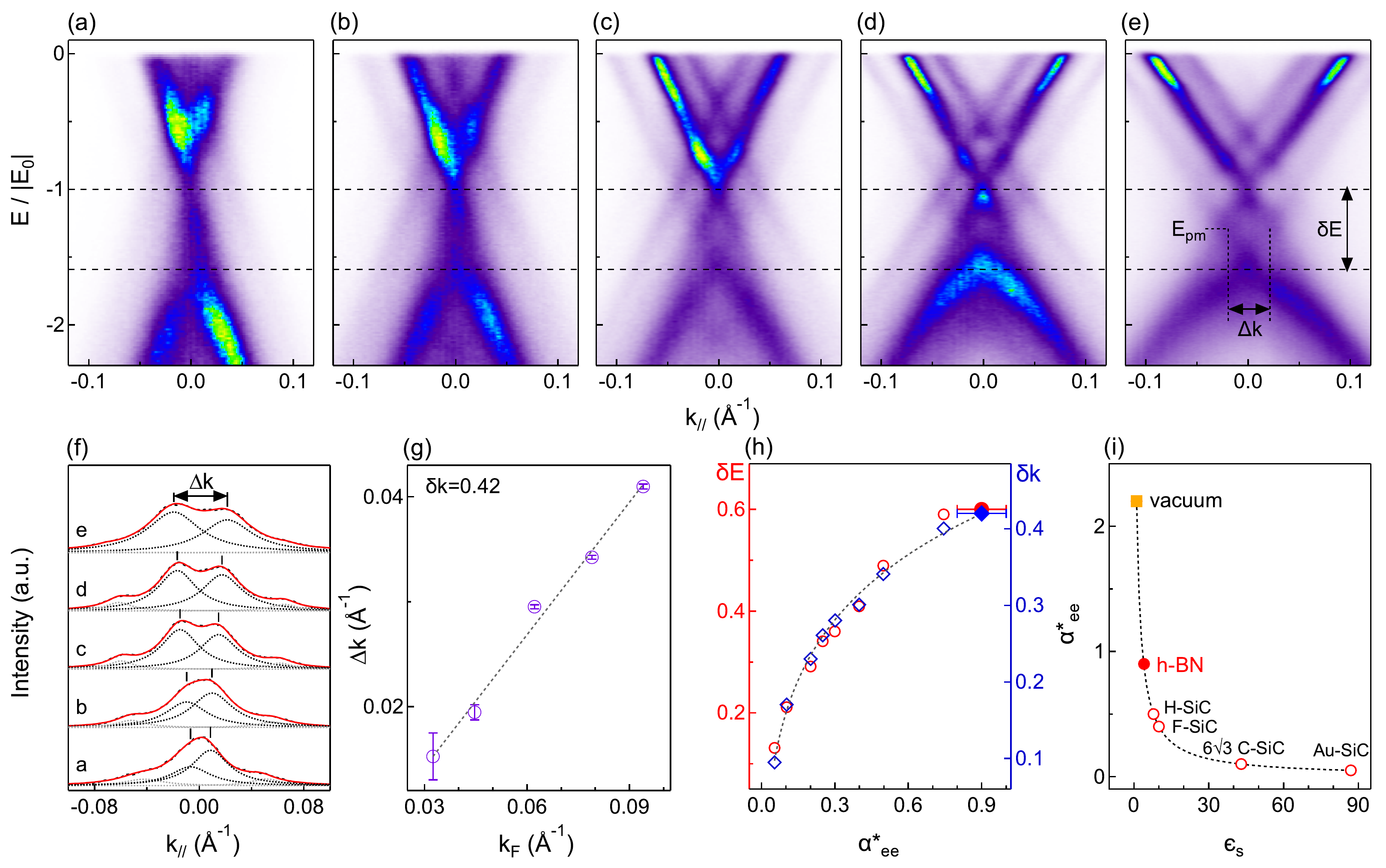}
	\caption{{\bf The extracted effective fine structure constant in graphene/h-BN heterostructure.} (a)-(e) Renormalized dispersions for data at high doping regime with energy scaled to $\vert$${\rm E_0}$$\vert$. $\delta E$ represents the extracted dimensionless energy separation of the diamond shape after renormalization. (f) MDCs at energy of ${\rm E_{pm}}$ for extracting the momentum separation $\Delta k$ of the diamond-shaped dispersion. (g) Extracted $\Delta k$ vs $\rm k_F$ with a slope of $\delta k$ = $\Delta k$/$\rm k_F$ = 0.42 $\pm$ 0.02. (h) $\delta E$ and $\delta k$ as a function of the dimensionless constant $\alpha_{ee}^*$. Filled symbols are extracted values for our graphene/h-BN heterostructure and open symbols are from theoretical calculations \cite{EliPRB2011} as a reference. The dotted curve is the fitting curve to infer the $\alpha_{ee}^*$ with larger $\delta E$ and $\delta k$ value. (i) Comparison of $\alpha_{ee}^*$ for our graphene/h-BN sample (red filled symbol), reported values for other graphene systems (open symbols),\cite{EliPRB2011} and theoretical limit for graphene in vacuum (orange filled symbol). Black dotted curve is a plot of $\alpha_{ee}^* = {\rm e^2}/\epsilon \hbar {\rm v_F}$ $\approx$ 4.4/($\epsilon_s$+1).}
	\label{fig5}
\end{figure*}

To quantify the energy and momentum separation between the plasmaron bands and the graphene Dirac cone, we show in Fig.~5 a detailed analysis of the plasmaron dispersion upon doping for 0$^\circ$ aligned graphene/h-BN. 
Figure~5a-e shows the renormalized dispersions with energy $E$ scaled to $\vert$${\rm E_0}$$\vert$, with ${\rm E_0}$ defined as the shift of the Dirac point from ${\rm E_F}$. In the renormalized plot, the diamond-shaped dispersion is positioned at the same energy with similar energy range $\delta E$ (indicated by broken lines) for all doping levels, confirming that the dimensionless energy separation $\delta E$ = $\vert$${\rm E_{1}}$ - ${\rm E_{0}}$$\vert$/$\vert$${\rm E_0}$$\vert$ = 0.60 $\pm$ 0.02 (${\rm E_0}$ and ${\rm E_1}$ are labeled in Fig.~4d) is doping independent. To quantify the size of the diamond-shaped dispersion along the momentum direction, we plot in Fig.~5f the MDCs at ${\rm E_{pm}}$ (labeled in Fig.~4d) with the fitting peaks appended. The momentum separation $\Delta k$ of the two main peaks defines the momentum range of the diamond-shaped dispersion. Figure~5g shows that the extracted momentum separation $\Delta k$ scales linearly with the Fermi momentum ${\rm k_F}$, and the slope gives a renormalized dimensionless momentum separation $\delta k$ = $\Delta k$/$\rm k_F$ = 0.42 $\pm$ 0.02. We note that the doping independent $\delta E$ and $\delta k$ are consistent with results calculated from the spectral function $A(k,w)$, and they are uniquely determined by the $\alpha_{ee}^*$ value.\cite{EliPRB2011} Figure~5h shows a replot of calculated $\delta E$ and $\delta k$ at different values of $\alpha_{ee}^*$ (open symbols).\cite{EliPRB2011} Our extracted $\delta E$ and $\delta k$ (filled symbols) fall on the extrapolated curves and correspond to $\alpha_{ee}^* \approx0.9\pm 0.1$.

Figure~5i shows a comparison of experimental $\alpha_{ee}^*$ values for graphene samples reported so far, including graphene samples grown on Au-, fluoride-, and hydrogen-treated SiC substrate as well as the carbon face of SiC.\cite{Eliplasmaron2010,EliPRB2011} Our graphene/h-BN heterostructure shows the largest reported $\alpha_{ee}^*$ value among all graphene samples. At the lowest order approximation, the effective fine structure constant is related to the dielectric environment by $\alpha_{ee}^* = {\rm e^2}/\epsilon \hbar {\rm v_F}$, which is directly determined by the dielectric environment. We note that similar spectral features  have also been discussed theoretically as satellite bands induced by weakly interacting electron-plasmon interaction\cite{Steven_PRL2013} rather than strongly interacting plasmarons, which corresponds to a much larger electron-electron interaction strength. Therefore, our extracted value of $\alpha_{ee}^* \approx0.9\pm 0.1$ gives the lower limit of the effective fine structure constant in graphene/h-BN heterostructure. Without considering the dielectric screening from the valence electrons and the Rb atoms, which is quite reasonable considering that $\delta E$ and $\delta k$ are both independent of carrier concentration or amount of Rb deposited, the effective dielectric constant is taken as the average value between dielectric constants of materials on both sides of graphene. 
In the extreme case for free-standing graphene without dielectric screening from the environment, $\epsilon =1$ and the effective fine structure constant is $\alpha_{ee}^* = 2.2$ (orange symbol in Fig.~5i), which is an upper limit for $\alpha_{ee}^*$.  
When graphene is placed on a substrate with dielectric constant $\epsilon_s$, the effective fine structure constant depends on the substrate dielectric constant $\epsilon_s$ by  $\alpha_{ee}^* \approx 4.4/(\epsilon_s+1)$\cite{EliPRB2011}, where the effective dielectric constant $\epsilon$ is taken as the average between the vacuum $\epsilon_{vac}$ = 1 and $\epsilon_s$. The fitting function of $\alpha_{ee}^* \approx 4.4/(\epsilon_s+1)$ is also plotted in Fig.~5i. From this relation, the extracted $\alpha_{ee}^* \approx 0.9$ for graphene/h-BN gives an effective dielectric constant of $\epsilon \approx 2.5$, which corresponds to substrate dielectric contribution $\epsilon_s \approx$ 4, similar to the reported dielectric constant of $\epsilon_{h-BN} \approx$ 3-4.\cite{YangNaturematerials2013,Hone2010} Considering the dielectric constant of h-BN, the large fine structure constant of graphene/h-BN is not surprising, however, being able to experimentally observing it is still an important experimental progress.

\section*{Discussions}
In summary, we report the experimental evidence of plasmaron and extract the fine structure constant of graphene/h-BN.  We note that experimental values for the effective fine structure constant have been reported through optical transparency\cite{Nair_Sciencec2008} and inelastic x-ray scattering measurement\cite{AbbamonteSci,Abbamonte_PRB2016} on both graphene and graphite, and the fitting of the Dirac point velocity,\cite{LanzaraPRL2013} which involves both effects of the carrier screening and dielectric screening. 
Here by observing not only the Dirac cone but also the dispersion of the previously inaccessible plasmarons in graphene/h-BN at high electron density, we extract the dressed effective fine structure constant  $\alpha_{ee}^* \approx 0.9$. Such large effective fine structure constant reveals the important  role of the small dielectric constant of h-BN in reducing the dielectric screening. 
In addition, the dispersions remain quite sharp (Fig.~2c-h) under the presence of such a large number of carriers and strong plasmon-charge interaction, suggesting the insignificant contribution of both scattering channels in the electron scattering.  Since many device applications require tunable electronic density, our finding on the small scattering of a highly electron-doped graphene/h-BN provides useful information for applications in gate-tunable nano-electronic and nano-plasmonic applications.\cite{BasovPlasmon2015,KoppensPlasmon2015}

In the past decade, h-BN has been widely used as a substrate and a capping layer, for example, in magic-angle twisted bilayer graphene or ABC stacking  trilayer graphene on h-BN, both exhibiting Mott insulator\cite{PabloMott2018,WangFTrilayerMott2019} and superconductivity upon doping,\cite{PabloSC2018,WangFTrilayerSC2019} Revealing the effect of dielectric property of h-BN on the effective fine structure constant of graphene can also be helpful for understanding the electron-electron interaction in graphene/h-BN. 
Finally, it has been suggested that the effective fine structure constant is relevant to other Dirac systems\cite{Abbamonte_PRB2016} including topological insulator surface states,\cite{Kane_RevModPhys2010,Zhang_RevModPhys2011} and Dirac or Weyl materials.\cite{Wan_PhysRevB2011,Young_PhysRevLett2012} Therefore, our results on graphene/h-BN heterostructure in principle can be extended to other Dirac materials for evaluating the electron-electron interaction and the effective fine structure constant.

\begin{methods}

\subsection{Sample preparation.}

Sample preparation of the $0^{\circ}$ aligned graphene/h-BN sample.	Single crystal h-BN flakes were first exfoliated onto a 300 nm SiO$_2$$/$Si substrate by mechanical cleaving method. Graphene samples were directly grown on h-BN substrates by the epitaxial method, as specified in previous work.\cite{YangNaturematerials2013} As-grown samples were characterized by tapping mode atomic force microscopy at room temperature in ambient atmosphere. We used freshly cleaved mica as shadow masks for metal electrode deposition. The contact metal (2 nm Cr on 90 nm Au) was deposited on the non-mica-covered area with a small part of target graphene$/$h-BN samples. The samples were then annealed at 200 $^\circ$C, after removing the mica flakes. 

Sample preparation of the $13.5^{\circ}$ twisted graphene/h-BN sample. Firstly, h-BN flake was exfoliated onto a PDMS stamp. Then the graphene flake on a SiO$_2$/Si substrate was picked up sequentially with h-BN on PDMS, and the twist angle between graphene and h-BN was determined by the edge of the flakes. Subsequently, the graphene/h-BN structure was flipped over and picked up with a second PDMS stamp and then transferred onto the gold-plated substrate. Finally, a piece of graphite was placed to connect graphene and gold to make sure the conductivity.

\subsection{ARPES measurements.}
ARPES measurements were performed at beamline 4.0.3 and 12.0.1 of the Advanced Light Source at Lawrence Berkeley National Laboratory (LBNL). The optimal spot size was set to 30 $\upmu$m. The data were recorded with photon energies of 50 and 60 eV. The overall energy and angle resolution are better than 26 meV and 0.1$^\circ$, respectively. Before measurements, the samples were annealed at 200-300$^\circ$C until sharp dispersions were observed. All measurements were performed below 20 K and under a vacuum better than 5$\times$$10^{-11}$ torr. The Rb deposition was achieved by heating an SAES commercial dispenser {\it in situ}.

\end{methods}

%% Here is the endmatter stuff: Supplementary Info, etc.
%% Use \item's to separate, default label is "Acknowledgements"

\begin{addendum}
 \item We thank Yuanbo Zhang, V. I. Fal'ko and J. Jung for useful discussions. This work is supported by National Key R $\&$ D Program of China (Grant No.~2016YFA0301004, 2020YFA0308800), National Natural Science Foundation of China (Grant No.~11725418 and 11427903), Beijing Advanced Innovation Center for Future Chip (ICFC), Tsinghua University Initiative Scientific Research Program and Tohoku-Tsinghua Collaborative Research Fund, Science Challenge Project (Grant No.~TZ2016004). 
K.W. and T.T. acknowledge support from the Elemental Strategy Initiative conducted by the MEXT, Japan, Grant Number JPMXP0112101001, JSPS KAKENHI Grant Number JP20H00354 and the CREST(JPMJCR15F3), JST.
This research used resources of the Advanced Light Source, which is a DOE Office of Science User Facility under contract No. DE-AC02-05CH11231.

\item[Competing Interests] 

The Authors declare no Competing Financial or Non-Financial Interests.

\item[Author Contributions] S.Z. designed the research project. Hongyun Z., E.W., C.B., K.D., H.Z., A.F., J.D. and S.Z. performed the ARPES measurements and analyzed the ARPES data. S.W., X.L., Q. L., G.C. and G.Z. prepared the graphene samples. K.W. and T.T. prepared h-BN crystals. Hongyun Z. and S.Z. wrote the manuscript, and all authors commented on the manuscript.

\end{addendum}

\end{document}